\documentclass[twocolumn,showpacs,preprintnumbers,amsmath,amssymb,superscriptaddress]{revtex4}

\usepackage{graphicx}
\usepackage{graphicx}
\begin{document}
\title{Differences in the high-energy kink between hole- and electron-doped high-$T_{\rm c}$ superconductors}

\author{M. Ikeda}
\affiliation{Department of Physics and Department of Complexity 
Science and Engineering, University of Tokyo, 
Hongo 7-3-1, Bunkyo-ku, Tokyo 113-0033, Japan}

\author{T. Yoshida}
\affiliation{Department of Physics and Department of Complexity 
Science and Engineering, University of Tokyo, 
Hongo 7-3-1, Bunkyo-ku, Tokyo 113-0033, Japan}

\author{A. Fujimori}
\affiliation{Department of Physics and Department of Complexity 
Science and Engineering, University of Tokyo, 
Hongo 7-3-1, Bunkyo-ku, Tokyo 113-0033, Japan}

\author{M. Kubota}
\affiliation{Institute of Material Structures Science, High Energy Accelerator Research Organization (KEK), 
Oho 1-1, Tsukuba, Ibaraki 305-0801, Japan}

\author{K. Ono}
\affiliation{Institute of Material Structures Science, High Energy Accelerator Research Organization (KEK), 
Oho 1-1, Tsukuba, Ibaraki 305-0801, Japan}

\author{Y. Kaga}
\affiliation{Department of Advanced Materials Science, University of Tokyo, 
Kashiwanoha 5-1-5, Kashiwa, Chiba 277-8561, Japan}

\author{T. Sasagawa}
\affiliation{Department of Advanced Materials Science, University of Tokyo, 
Kashiwanoha 5-1-5, Kashiwa, Chiba 277-8561, Japan}
\affiliation{Materials and Structures Laboratory, Tokyo Institute of Technology, 
Nagatsuta 4259, Midori-ku, Yokohama, Kanagawa 226-8503, Japan}

\author{H. Takagi}
\affiliation{Department of Advanced Materials Science, University of Tokyo, 
Kashiwanoha 5-1-5, Kashiwa, Chiba 277-8561, Japan}

\date{\today}
\begin{abstract}

We have performed an angle-resolved photoemission spectroscopy (ARPES) study of Nd$_{1.85}$Ce$_{0.15}$CuO$_{4}$ (NCCO) in order to elucidate the origin of the high-energy kink (HEK) observed in the high-$T_{\rm c}$ superconductors (HTSCs). The energy scale of the HEK in NCCO is large compared with that in hole-doped HTSCs, consistent with previous ARPES studies. From measurement in a wide momentum region, we have demonstrated that between the hole- and electron-doped HTSCs the energy position of the HEK is shifted approximately by the amount of the chemical potential difference. Also, we have found that around $(\pi, 0)$ the HEK nearly coincides with the band bottom while around the node the band reaches the incoherent region and the HEK appears at the boundary between the coherent and incoherent regions.

\end{abstract}
\maketitle

\section{Introduction}

Various unusual electronic structures of the high-$T_{\rm c}$ superconductors (HTSCs) have so far been demonstrated by angle-resolved photoemission spectroscopy (ARPES) studies. Among them, an anomaly of the band dispersion in the high-energy region of 0.3-0.5 eV has recently been identified experimentally and debated from experimental \cite{Ronning2005, Graf2007, Graf20072, Xie2007, Valla2007, Chang2007, Inosov2007, Inosov2008, Pan, Meevasana2007, Zhang2008, Hwang2007, Meevasana2008} and theoretical points of view \cite{Macridin2007, Markiewicz2007, Srivastava2007, Byczuk2007, Zemljic2008, Tan2008, Zhou2007, Alexandrov2007, Cojocaru2007, Kim2007, Tan2007, Markiewicz20072, Zhu2008, Leigh2007, Manousakis2007}. The anomaly is called high-energy kink (HEK) or ``waterfall'' and has been interpreted in different ways: Disintegration of the quasiparticle into a spinon and a holon \cite{Graf2007, Graf20072}, polaronic effects \cite{Xie2007}, high-energy spin fluctuations \cite{Valla2007, Macridin2007, Markiewicz2007, Srivastava2007}, coherence-incoherence crossover \cite{Chang2007, Byczuk2007}, and ARPES matrix element effects \cite{Inosov2007, Inosov2008}. However, most of the above interpretations of the HEK are based on the results of hole-doped HTSCs. In order to elucidate the mechanism of the HEK, it is necessary to look into electron-doped HTSCs, too, and to identify the differences and similarities between the hole- and electron-doped HTSCs. Theoretical studies using the $t$-$J$ model have demonstrated that there is no HEK in the electron-doped HTSCs due to the lack of the incoherent part in the photoemission spectra \cite{Zemljic2008, Tan2008}, while other theoretical studies have demonstrated that a HEK of the electron-doped HTSCs exists in a high-energy region compared with that of the hole-doped ones due to a charge modulation mechanism \cite{Zhou2007} or due to the different magnetic susceptibilities of the hole- and electron-doped HTSCs according to a paramagnon-induced HEK mechanism \cite{Markiewicz2007}. Since most of the studies of the HEK have been performed in a limited region of momentum space, more systematic investigations of the momentum dependence of the HEK are necessary to understand the origin of the HEK.

In this work, we report on an ARPES study of the prototypical electron-doped HTSC Nd$_{1.85}$Ce$_{0.15}$CuO$_{4}$ in a large energy-momentum space and compare the results with those of the hole-doped ones. The observed energy scale of the HEK was large compared with that in the hole-doped HTSCs, consistent with the previous ARPES studies \cite{Pan}. The momentum dependence of the HEK indicates that the difference of the HEK between the hole- and electron-doped HTSCs can be largely ascribed to that of the chemical potential. Also, different origins of the HEK in different momentum regions shall be discussed.

\begin{figure*}
\begin{center}
\includegraphics[width=16cm]{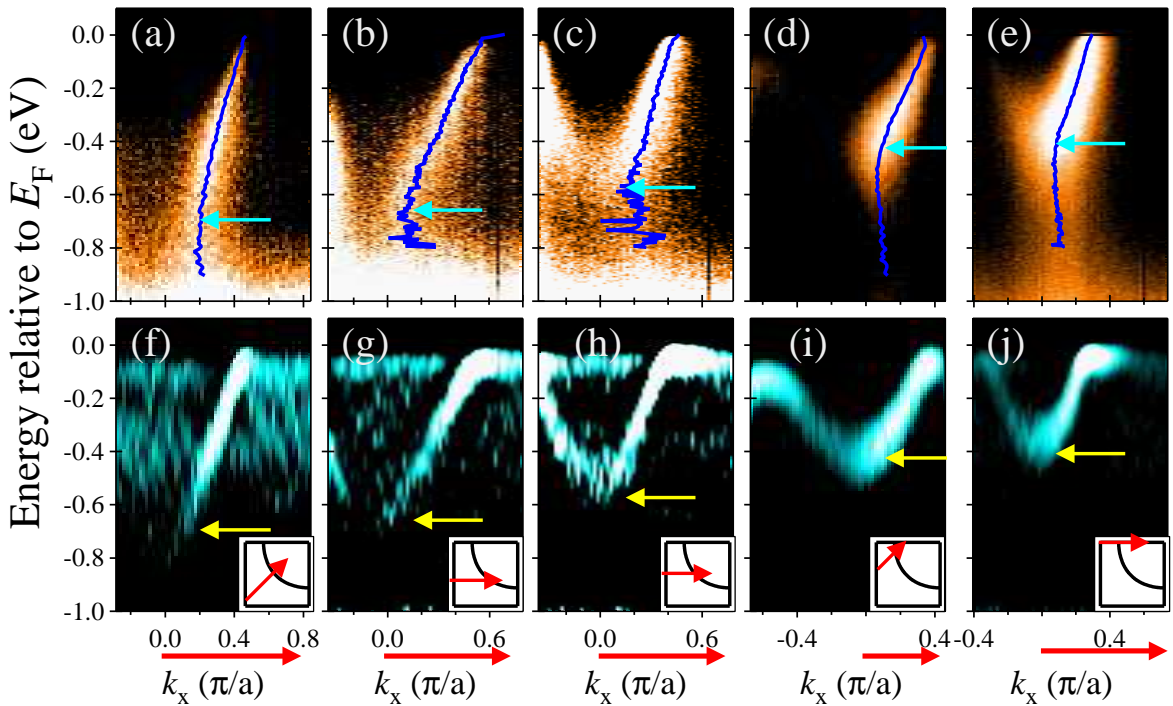}
\caption{(Color online) Plots of ARPES intensities in energy-momentum space for NCCO along various cuts shown in the insets. (a)-(e) Raw data. (f)-(j) Corresponding second derivatives of energy distribution curves. Data in (a), (d) are taken with $h\nu= 100$ eV, and those in (b), (c), and (e) are taken with $h\nu=55$ eV. A horizontal arrow in each panel represents the position of the HEK. In the upper panels, the peak positions of momentum distribution curves are shown to represent the band dispersion.}
\end{center}
\end{figure*}

\section{Experiment}

High-quality single crystals of optimally doped Nd$_{1.85}$Ce$_{0.15}$CuO$_{4}$ (NCCO) were grown by the traveling solvent floating zone method. Single crystals of NCCO were annealed at 920 $^{\circ}$C for 24 hours in Ar gas. The $T_{\rm c}$ of NCCO was $\sim$22 K. The ARPES measurements were performed at beamline 28A of Photon Factory (PF), Institute of Materials Structure Science, High Energy Accelerators Research Organization (KEK), using circularly-polarized light with energies of 55 eV and 100 eV. We used a SCIENTA SES-2002 electron-energy analyzer and a five-axis manipulator \cite{Aiura2003}. The total energy resolution and angular resolution were 15-60 meV and 0.2$^\circ$, respectively. Samples were cleaved {\it in situ} in an ultrahigh vacuum of 10$^{-9}$ Pa. The incident angle of the photon beam was approximately 45$^\circ$ to the sample surface. We made the ARPES measurements at $\sim$10 K. The Fermi edge of gold was used to determine the Fermi level ($E_{\rm F}$) position and the instrumental resolution before and after the ARPES measurements. The spectral intensities have been normalized to the intensity above $E_{\rm F}$, which arises from the second order light of the monochromator.

\section{Results and Discussion}

Figure 1 shows ARPES intensity plots of NCCO in energy-momentum space. The momentum cut is shown in the insets of each panel. In the nodal region, in the previous ARPES studies of hole-doped HTSCs, a HEK was observed around 0.3-0.5 eV. In NCCO, a HEK is seen at a larger binding energy (0.6-0.8 eV) [Fig. 1(a)]. This large energy range of the HEK is close to that in Pr$_{1-x}$LaCe$_{x}$CuO$_{4}$ (PLCCO) \cite{Pan}, indicating a common energy scale for the electron-doped HTSCs. In Figs. 1(b)-(e), ARPES intensity plots along other cuts are shown. In going from the nodal region [Fig. 1(a)] to the $(\pi, 0)$ region [Fig. 1(e)], one can see that the binding energy of the HEK becomes small as indicated by arrows in the upper panels of Fig. 1. This momentum dependence of the HEK in NCCO is qualitatively similar to that in the previous ARPES studies on Bi$_{2}$Sr$_{2}$CaCu$_{2}$O$_{8+ \delta}$ (Bi2212) \cite{Graf20072, Zhang2008}, La$_{2-x}$Ba$_{x}$CuO$_{4}$ \cite{Valla2007}, and La$_{2-x}$Sr$_{x}$CuO$_{4}$ (LSCO) \cite{Chang2007}, indicating a qualitative similarity of the HEK between the hole- and electron-doped HTSCs, except for the overall energy position of the HEK.

Let us discuss the momentum dependence of the HEK in more detail. In the lower panels of Fig. 1, we present the second derivatives of energy distribution curves of the ARPES intensity plots shown in the upper panels. Along the cut across the $(\pi, 0)$ region [Fig. 1(e) and (j)], the band crossing $E_{\rm F}$ at $k_{x}$ $\sim -0.2$ $\pi/a$ and that crossing $E_{\rm F}$ at $k_{x}$ $\sim 0.2$ $\pi/a$ merge around $-0.4$ eV while along the cut across the $(3\pi/4, 0)$ region [Fig. 1(b) and (g)] the bands do not merge down to $-0.7$ eV, suggesting two kinds of HEK depending on the momentum. As for the former HEK, the energy position [arrow in Fig. 1(e)] approximately coincides with the band bottom [arrow in Fig. 1(j)], and below $-0.4$ eV the vertical dispersion of the momentum distribution curve (MDC) peak appears due to the intensity tail, indicating that the HEK around $(\pi, 0)$ occurs near the band bottom. The incoherent part may be located well below the band bottom. As for the latter HEK around the node, one can see that the band disappears and MDC-peak dispersion becomes vertical before arriving at the band bottom, probably entering the incoherent regime.

The HEK positions, namely, the position of the ``vertical'' dispersion, in two-dimensional momentum space are plotted in Fig. 2. In NCCO, around ($\pi, 0$), the HEK positions are close to the ($0, 0$)-($\pi, 0$) line, where the band bottom is located. Here, the slight deviation of the HEK position from the zone boundary may partly be due to the asymmetric intensity distribution caused by the circularly polarized light. On the other hand, around the node the HEK positions strongly deviate from this line, suggesting that the HEK positions do not correspond to the band bottom. This tendency has also been observed for LSCO as shown in Fig. 2(b) \cite{Chang2007}. It is interesting to note that the HEK positions are similar between NCCO and LSCO in spite of the very different Fermi surfaces. This suggests that the different chemical potential position in the band structure between NCCO and LSCO do not affect the behavior of the HEK as we shall see below. Here, one cannot exclude the possibility that the effects of matrix elements are the origin of the HEK around the node \cite{Inosov2007, Inosov2008}. However, although we performed the experiment in two kinds of geometries and photon energies, significant difference was not observed in the band dispersion. Therefore, we consider that from the $(\pi, 0)$ region to $\sim (\pi/2, 0)$, the HEK occurs near the band bottom while from $\sim (\pi/2, 0)$ to the nodal region, it is due to the boundary between the coherent and incoherent regions before it reaches the band bottom.

\begin{figure}
\begin{center}
\includegraphics[width=7cm]{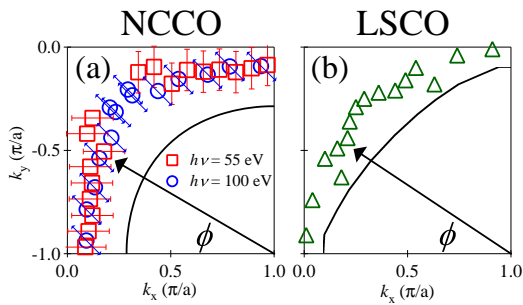}
\caption{(Color online) HEK positions in two-dimensional momentum space for NCCO (a) and LSCO (b). NCCO data points have been taken using $h\nu=55$ eV and $h\nu=100$ eV. LSCO data points have been reproduced from Ref. \cite{Chang2007}. Solid curve is the Fermi surface. The data were taken over a Brillouin zone octant and symmetrized with respect to the (0,0)-$(\pi, \pi)$ line.}
\end{center}
\end{figure}

In Fig. 3(a), we show the HEK positions of NCCO, together with those of LSCO \cite{Chang2007}, La$_{1.64}$Eu$_{0.2}$Sr$_{0.16}$CuO$_{4}$ (Eu-LSCO) \cite{Graf20072}, and Bi2212 \cite{Graf20072}. In order to quantitatively examine the momentum dependence of the HEK, we have fitted the $d$-wave order parameter to the energy position of the HEK, as suggested in Ref. \cite{Chang2007}, where the authors have found a $d$-wave-like behavior for the HEK and suggested that the HEK may be related to the superconducting gap. Here, the fitted $d$-wave-like gap function is given by $A(1-|$cos$(2 \phi)|)+B$ where $A=-0.36$ and $B=-0.39$ for NCCO, as shown in Fig. 3(a). The fitted results indicate that the major difference between LSCO ($A=-0.43$, $B=0$) and NCCO ($A=-0.36$, $B=-0.39$) is the value of $B$, that is, the position of the HEK in NCCO is rather uniformly shifted downward by $\sim$0.4 eV relative to the position of the HEK in LSCO. This can be understood as due to the difference in the chemical potential in the nearly common band structures of NCCO and LSCO. Note that although whether the difference of $\sim$0.4 eV corresponds to the real chemical potential difference between NCCO and LSCO is under dispute due to the different crystal structure, the uniform shift of $\sim$0.4 eV between NCCO and LSCO was estimated from the core-level X-ray photoemission studies \cite{Ino1997, Harima2001}.

\begin{figure}
\begin{center}
\includegraphics[width=9cm]{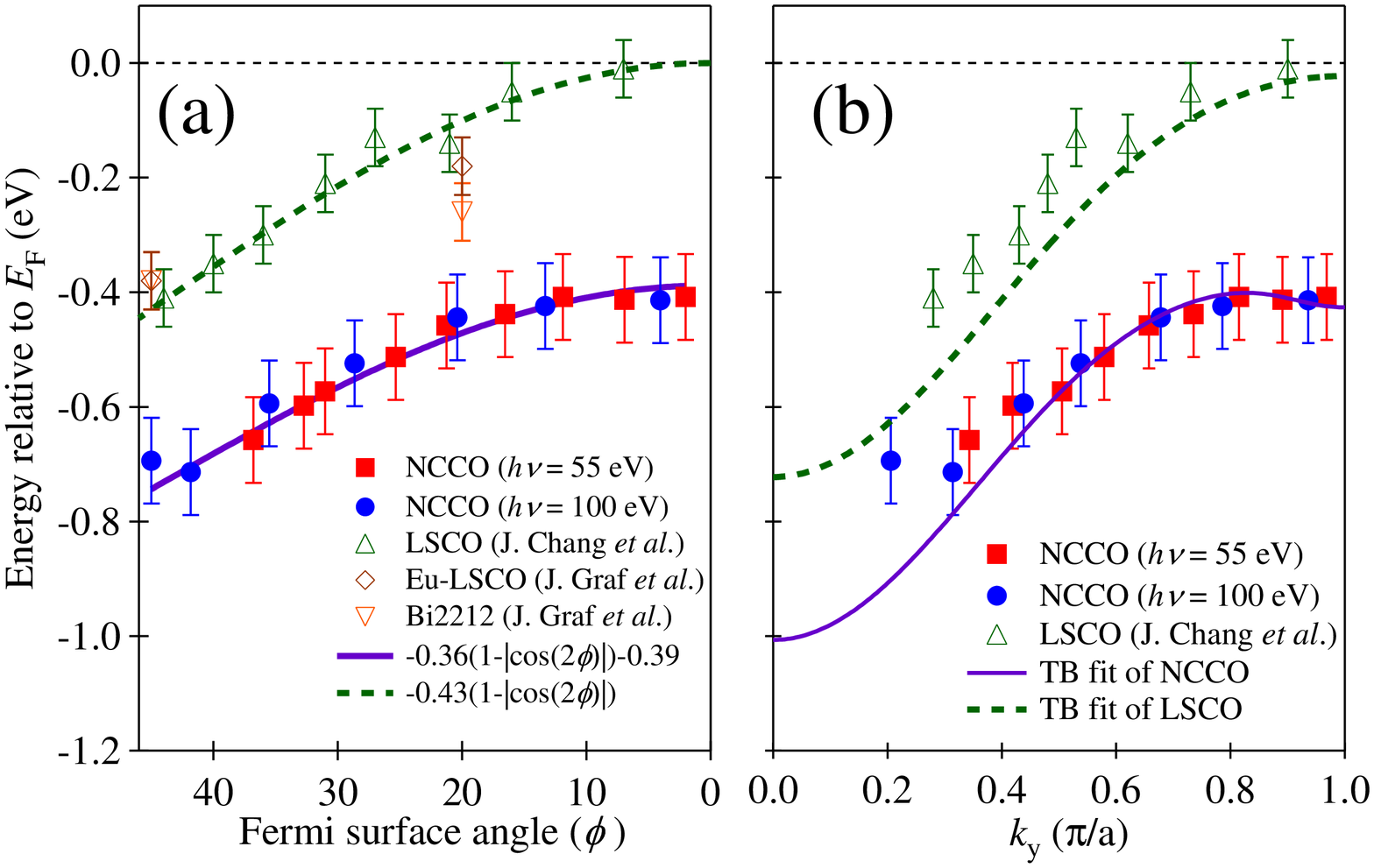}
\caption{(Color online) Positions of the HEK for NCCO (squares and circles), together with LSCO (triangles) \cite{Chang2007}, La$_{1.64}$Eu$_{0.2}$Sr$_{0.16}$CuO$_{4}$ (Eu-LSCO) (diamonds) \cite{Graf20072}, and Bi2212 (inverse triangles) \cite{Graf20072}. The definition of the Fermi surface angle is depicted in Fig. 2(a). In going from the ($\pi, 0$) region to the nodal region, the HEK position moves towards high binding energies. (a) The solid curve is a $d$-wave-gap-like function fitted to the HEK positions for NCCO. The dashed curve is that for LSCO \cite{Chang2007}. (b) Comparison with the tight-binding (TB) model. TB parameters for NCCO and LSCO are taken from Refs. \cite{Ikeda} and \cite{Yoshida2006}, respectively.}
\end{center}
\end{figure}

In Fig. 3(b), we compare the HEK positions with the band bottom estimated from the tight-binding (TB) model:
\begin{eqnarray}
\epsilon - \mu= \varepsilon_{0} - \sqrt{\Delta E^{2}+4t^{2}(\cos{k_{x}a}+\cos{k_{y}a})^{2}}\nonumber \\-4t'\cos{k_{x}a} \cos{k_{y}a}-2t''(\cos{2k_{x}a}+\cos{2k_{y}a}) \nonumber,
\end{eqnarray}
where $t$, $t'$, and $t''$ are the transfer integrals between the nearest-neighbor, second-nearest-neighbor, and third-nearest-neighbor Cu sites, respectively, $\varepsilon_{0}$ represents the center of the band relative to the chemical potential $\mu$, and 2$\Delta E$ is the potential energy difference between the spin-up and spin-down sublattices. The TB parameters for NCCO ($t=0.27$, $-t'/t=0.20$, $-t''/t'=0.5$, $\Delta E=0.07$, and $\varepsilon_{0}/t=-0.12$), and those for LSCO ($t=0.25$, $-t'/t=0.15$, $-t''/t'=0.5$, $\Delta E=0$, and $\varepsilon_{0}/t=0.81$), are taken from Refs. \cite{Ikeda} and \cite{Yoshida2006}, respectively. Around ($\pi, 0$), the HEK positions are well fitted by the band bottom of the TB model while around the node there is a deviation between the HEK and the band bottom, consistent with the above discussion.

Finally, we discuss the material dependence of the HEK position. As shown in Fig. 3(a), while around the node the position of the HEK is similar among the different hole-doped HTSCs, namely, among LSCO, Eu-LSCO, and Bi2212, around $\phi=20 ^\circ$ the binding energy of the HEK in Bi2212 is a little larger than those in LSCO and Eu-LSCO. The difference among the materials can be explained by the different position of the ``($\pi, 0$) flat band". The position of the flat band is strongly dependent on the value of $-t'$. According to the previous studies \cite{Tanaka2004, Pavarini2001, Tohyama2000}, $-t'$ of LSCO is smaller than that of Bi2212. Therefore, we consider that the difference in $-t'$ among them leads to the difference of the HEK positions, confirming that the origin in the HEK around ($\pi, 0$) results from the band bottom.

\section{Conclusion}

In conclusion, we have performed an ARPES study of NCCO in order to elucidate the origin of the HEK observed in HTSCs. The energy of the HEK in NCCO is large compared with that in the hole-doped HTSCs, similar to that in PLCCO \cite{Pan}. The present measurements in a wide momentum range elucidated that the difference of the HEK between the hole- and electron-doped HTSCs is largely attributed to the difference in the chemical potential. The momentum dependence of the HEK demonstrates that, around the $(\pi, 0)$ region, the HEK occurs near the band bottom while in the nodal region the HEK is related to the end point of the band dispersion possibly due to the boundary between the coherent and incoherent parts.

\section*{ACKNOWLEDGMENTS}

We are grateful to H. Ding and T. Tohyama for fruitful discussions, and N. Kamakura and Y. Kotani for technical support at beamline 28A of Photon Factory. This work was done under the approval of the Photon Factory Program Advisory Committee (Proposal No. 2006S2-001), and was supported by a Grant-in-Aid for Scientific Research in Priority Area ``Invention of Anomalous Quantum Materials'' from the Ministry of Education, Culture, Sports, Science and Technology (MEXT), Japan. We also thank the Material Design and Characterization Laboratory, Institute for Solid State Physics, University of Tokyo, for the use of a SQUID magnetometer. This work was supported in part by Global COE Program, MEXT, Japan.

\end{document}